\documentclass[10pt,aps,twocolumn,pra,superscriptaddress]{revtex4-1}
\usepackage{amsfonts}

\usepackage{amsmath,amssymb}
\usepackage{graphicx,color}
\usepackage{epstopdf}
\usepackage{epsfig}
\usepackage{mathrsfs}
\usepackage[breaklinks=true]{hyperref}

\newcommand{\beq}{\begin{equation}}
\newcommand{\eeq}{\end{equation}}
\newcommand{\bqa}{\begin{eqnarray}}
\newcommand{\eqa}{\end{eqnarray}}

\definecolor{green}{rgb}{0.00,0.50,0.00}

\newtheorem{theorem}{Theorem}

\newtheorem{proposition}[theorem]{Proposition}
\newtheorem{remark}[theorem]{Remark}

\newenvironment{proof}[1][Proof]{\noindent\textbf{#1.} }{\ \rule{0.5em}{0.5em}}

\begin{document}

\title{Isolated Loops in Quantum Feedback Networks}

\author{John E.~Gough} \email{jug@aber.ac.uk}
   \affiliation{Aberystwyth University, SY23 3BZ, Wales, United Kingdom}
\author{Symeon Grivopoulos} \email{symeon.grivopoulos@gmail.com}
	 \affiliation{UNSW Canberra, Canberra BC 2610, Australia}
\author{Ian R. Petersen} \email{i.r.petersen@gmail.com}
	 \affiliation{Australian National University, Canberra, Canberra BC 2610, Australia}

\date{\today}

\begin{abstract}

A scheme making use of an isolated feedback loop was recently proposed in \cite{GP_} for creating an arbitrary bilinear Hamiltonian interaction between two multi-mode Linear Quantum Stochastic Systems (LQSSs). In this work we examine the presence of an isolated feedback loop in a general SLH network, and derive the modified Hamiltonian of the network due to the presence of the loop. In the case of a bipartite network with an isolated loop running through both parts, this results in modified Hamiltonians for each subnetwork, as well as a Hamiltonian interaction between them. As in the LQSS case, by engineering appropriate ports in each subnetwork, we may create desired interactions between them. Examples are provided that illustrate the general theory.
\end{abstract}

\maketitle

\section{Introduction}

In the theory of Markovian quantum-optical open systems with an explicit input-output structure, the SLH formalism \cite{GJ_QFN,GJ_Series} provides a fairly general modeling framework. Moreover, under the assumption that the propagation time of light from one lumped open system to another is very short, compared to the timescales of the internal dynamics of the systems involved, the SLH formalism provides a natural and efficient framework to model networks of such systems. The quantum network composition rules developed in \cite{GJ_QFN,GJ_Series}, see also \cite{CKS}, lead to explicit, tractable models of such networks.

An especially interesting subclass of SLH models, is given by Linear Quantum Stochastic Systems \cite{pet10} - \cite{GJN_squeeze}. Here, the system consists of several bosonic degrees of freedom (modes), and the coupling to the input fields is such that, both the ensuing Langevin equations and the input-output relations are affine linear in the mode. Besides their frequent use in quantum optics \cite{garzol00,walmil08,wismil10}, they also appear in circuit QED systems \cite{matjirper11,kerandku13}, and quantum opto-mechanical systems \cite{tsacav10,masheipir11,donfiokuz12}. Important proposals for LQSSs include their usage as coherent quantum feedback controllers for other quantum systems, i.e. controllers that do not perform any measurement on the controlled quantum system, and thus, have the potential to outperform classical controllers, see e.g. \cite{YK1} - \cite{critezsoh13}. The network composition rules have been used in the context of LQSSs for coherent quantum controller synthesis \cite{nurjampet09,JNP_2008}, for the synthesis of larger scale LQSSs \cite{nurjamdoh09} - \cite{gripet15}, etc.

In the context of LQSSs, a number of works has employed bilinear Hamiltonian interactions \cite{nurjamdoh09}, \cite{zhajam11} - \cite{sicvlapet15} in applications such as the synthesis of larger LQSSs in terms of simple ones, and the design of coherent quantum observers and controllers for LQSSs. Such interactions occur, for example, when light beams from different optical devices meet inside another optical device or material \cite{nurjamdoh09}. Except for a handful of case-specific implementations involving single-mode LQSSs \cite{nurjamdoh09,pethun15,pethun16b}, there does not exist a general scheme for the implementation of arbitrary bilinear Hamiltonian interactions between multi-mode LQSSs. Such a scheme was proposed recently in \cite{GP_}. It makes use of an isolated feedback loop between two LQSSs, that is a feedback loop where the inputs and outputs of every port used carry only loop signals (i.e. no external signal is injected into the loop, and no loop signal is output). It turns out that such a feedback loop creates a bilinear Hamiltonian interaction between the LQSSs, as well as modifies their internal Hamiltonians.

In this paper we study isolated loops in general quantum feedback networks.  We show in general, how such isolated loops modify the original network Hamiltonian. In the case of a bipartite network with an isolated loop running through both parts, this results in modified Hamiltonians for each subnetwork, as well as a Hamiltonian interaction between them. As in the LQSS case, by engineering appropriate ports in each subnetwork, we may create desired interactions between them.

The rest of the paper is organized, as follows: In Section \ref{sec:SLH}, we recall the basic framework, sometimes referred to as the $SLH$ formalism. In Section \ref{sec:Loops} we look at models with well-posed isolated loops in the setting of the general theory. In Section \ref{sec:Examples}, we look at some examples demonstrating the general theory. Section \ref{sec:Delay} examines the effect of time delay in a linear feedback network, and Section \ref{sec:Conclusions} concludes.

\section{The SLH Formalism}
\label{sec:SLH}

\subsection{Notation}

Let $\mathfrak{A}$ be a fixed algebra of operators, and let $M_{n,m}\left( \mathfrak{A}\right) $ denote the space of $n\times m$ arrays with entries in $\mathfrak{A}$. We say that $X=\left[ X_{ij}\right] \in M_{n,m}\left( \mathfrak{A}\right) $
and $Y=\left[ Y_{jk}\right] \in M_{m,r}\left( \mathfrak{A}\right) $ are
composable, and define their product $XY\in M_{n,r}\left( \mathfrak{A}\right) $
via the row into column law of matrix multiplication - i.e., the $ik$ entry
of $XY$ is $\sum_{j=1}^{r}X_{ij}Y_{jk}$.

In practice, $\mathfrak{A}$ will be a concrete algebra of operators on a fixed
Hilbert space $\mathfrak{h}$ in which case elements of $M_{n,m}\left( \mathfrak{A}%
\right) $ may be thought of as operators from $\oplus ^{m}\mathfrak{h}$ to $%
\oplus ^{n}\mathfrak{h}$. Frequently, we will label rows and columns by index
sets \textsf{i} and \textsf{j} and write $M_{\mathsf{ij}}\left( \mathfrak{A}%
\right) $ for the corresponding collection of arrays, and $\mathfrak{h}_{\mathsf{%
k}}=\oplus _{\mathsf{k}}\mathfrak{h}$. For instance, if we have a fixed index
set $\mathsf{k}$ which is a disjoint union of $\mathsf{i}$ and $\mathsf{j}$,
then we have the decomposition $\mathfrak{h}_{\mathsf{k}}\cong \mathfrak{h}_{\mathsf{%
i}}\oplus \mathfrak{h}_{\mathsf{j}}$, and for $X_{\mathsf{kk}}\in M_{\mathsf{kk}%
}\left( \mathfrak{A}\right) $ we have the corresponding block decomposition
\begin{eqnarray*}
X_{\mathsf{kk}}=\left[
\begin{array}{cc}
X_{\mathsf{ii}} & X_{\mathsf{ij}} \\
X_{\mathsf{ji}} & X_{\mathsf{jj}}
\end{array}
\right]
\end{eqnarray*}
with the sub-block $X_{\mathsf{ii}}$ belonging to $M_{\mathsf{ii}}\left(
\mathfrak{A}\right) $, etc. Normally the index set $\mathsf{k}$ will be $%
\{1,2,\cdots ,n\}$, however we have occasion to include an index 0. We will
use notation such as $X_{0\mathsf{k}}\in M_{0\mathsf{k}}\left( \mathfrak{A}%
\right) $, for instance, to denote a row array $[X_{01},\cdots X_{0n}]$, etc.

Given a decomposition $X=\left[
\begin{array}{cc}
X_{\mathsf{aa}} & X_{\mathsf{ab}} \\
X_{\mathsf{ba}} & X_{\mathsf{bb}}
\end{array}
\right] $, we define the Schur complement of $X$ to be
\begin{eqnarray*}
\underset{\mathsf{b}}{\mathrm{Schur}}X\triangleq X_{\mathsf{aa}}-X_{\mathsf{%
ab}}X_{\mathsf{bb}}^{-1}X_{\mathsf{ba}}
\end{eqnarray*}
where we shall always assume that $X_{\mathsf{bb}}$ is invertible as an
operator on $\mathfrak{h}\otimes \mathbb{C}^{\mathsf{b}}$. Specifically, we
say that this is the Schur complement of $X$ obtained by shortening the set
of indices $\mathsf{b}$.

A key property that we shall use is that the order in which successive
shortening of indices are applied is not important. In particular,
\begin{eqnarray*}
\underset{\mathsf{b}_{1}\cup \cdots \cup \mathsf{b}_{n}}{\mathrm{Schur}}\,X=%
\underset{\mathsf{b}_{1}}{\mathrm{Schur}}\cdots \underset{\mathsf{b}_{n}}{%
\mathrm{Schur}}X
\end{eqnarray*}
for any disjoint sets $\mathsf{b}_{1},\cdots ,\mathsf{b}_{n}\subset \mathsf{j}$.

\subsection{Quantum Markov Models}

In the Markovian model of an open quantum system we consider a fixed Hilbert
space $\mathfrak{h}_{0}$ for the system and a collection of independent
quantum white noises $b_{k}\left( t\right) $ labeled by $k$ belonging to
some discrete set $\mathsf{k}=\left\{ 1,\cdots ,n\right\} $. That is, we
have
\begin{eqnarray*}
\left[ b_{j}\left( t\right) ,b_{k}\left( s\right) ^{\ast }\right] =\delta
_{jk}\delta \left( t-s\right) .
\end{eqnarray*}
The Schr\"{o}dinger equation is
\begin{eqnarray}
\dot{U}\left( t\right) =-i\Upsilon \left( t\right) \,U\left( t\right)
\label{eq:Schro}
\end{eqnarray}
where the stochastic Hamiltonian takes the form
\begin{eqnarray}
\Upsilon \left( t\right) &=& E_{00}+\sum_{j\in \mathsf{k}}E_{j0}b_{j}\left(
t\right) ^{\ast }+\sum_{k\in \mathsf{k}}E_{0k}\ b_{k}\left( t\right) \nonumber \\
&+&\sum_{j,k\in \mathsf{k}}E_{jk}b_{j}\left( t\right) ^{\ast }b_{k}\left(
t\right) .
\end{eqnarray}
Here we assume that the $E_{\alpha \beta }$ are operators on $\mathfrak{h}%
_{0}$ with $E_{\alpha \beta }^{\ast }=E_{\beta \alpha }$. (In the course of
this paper we will ignore technicalities related to unbounded operators.) We may write
\begin{eqnarray*}
\Upsilon \left( t\right) =\left.
\begin{array}{c}
\left[ 1,b_{\mathsf{k}}\left( t\right) ^{\dag }\right] \\
\quad
\end{array}
\right. \mathbf{E}\left[
\begin{array}{c}
1 \\
b_{\mathsf{k}}\left( t\right)
\end{array}
\right],
\end{eqnarray*}
where
\begin{eqnarray}
b_{\mathsf{k}}\left( t\right) =\left[
\begin{array}{c}
b_{1}\left( t\right) \\
\vdots \\
b_{n}\left( t\right)
\end{array}
\right] ,\quad \mathbf{E}=\left[
\begin{array}{cc}
E_{00} & E_{0\mathsf{k}} \\
E_{\mathsf{k}0} & E_{\mathsf{kk}}
\end{array}
\right],
\end{eqnarray}
with $E_{0\mathsf{k}}=\left[ E_{01},\cdots ,E_{0n}\right] $, $E_{\mathsf{k}%
0}=E_{0\mathsf{k}}^{\dag }$ and $E_{\mathsf{kk}}$ the $n\times n$ matrix
with entries $E_{jk}$ with $j,k\in \mathsf{k}$.

We may integrate the noises to get the following fundamental quantum
stochastic processes:
\begin{eqnarray}
B_{k}\left( t\right) &=&\int_{0}^{t}b_{k}\left( s\right) ds, \\
B_{j}\left( t\right) ^{\ast } &=&\int_{0}^{t}b_{j}\left( s\right) ^{\ast }ds,
\\
\Lambda _{jk}\left( t\right) &=&\int_{0}^{t}b_{j}\left( s\right) ^{\ast
}b_{k}\left( s\right) ds.
\end{eqnarray}

The Schr\"{o}dinger equation (\ref{eq:Schro}) is interpreted as the
Stratonovich quantum stochastic differential equation
\begin{gather}
dU\left( t\right) =-i\bigg\{E_{00} \otimes dt+\sum_{j\in \mathsf{k}%
}E_{j0}\otimes dB_{j}\left( t\right) ^{\ast }  \nonumber \\
+\sum_{k\in \mathsf{k}}E_{0k}\otimes dB_{k}\left( t\right) +\sum_{j,k\in
\mathsf{k}}E_{jk}\otimes d\Lambda _{jk}\left( t\right) \bigg\}\circ U\left(
t\right) ,
\end{gather}
which may be readily converted into the quantum It\={o} form of Hudson and
Parthasarathy. (In fact the latter is accomplished by Wick ordering the
noise fields $b_{k}(t)$ and $b_{k}^{\ast }(t)$ in (\ref{eq:Schro})\cite{G_Wong-Zakai}.)
The Stratonovich differentials may be algebraically defined as $X\left(
t\right) \circ dY\left( t\right) =X\left( t\right) dY\left( t\right) +\frac{1%
}{2}dX\left( t\right) .dY\left( t\right) $.

We obtain the general form of the constant operator coefficient quantum
stochastic differential equation for an adapted unitary process $U(t)$:
\begin{gather}
dU\left( t\right) =\bigg\{-\left( \frac{1}{2}L_{\mathsf{k}}^{\ast }L_{%
\mathsf{k}}+iH\right) \otimes dt  \nonumber \\
+\sum_{j\in \mathsf{k}}L_{j}\otimes dB_{j}\left( t\right) ^{\ast
}-\sum_{j,k\in \mathsf{k}}S_{jk}L_{k}^{\ast }\otimes dB_{k}\left( t\right)
\nonumber \\
+\sum_{j,k\in \mathsf{k}}(S_{jk}-\delta _{jk})\otimes d\Lambda _{jk}\left(
t\right) \bigg\}U\left( t\right),
\end{gather}
where the $S_{jk},L_{j}$ and $H$ are operators on the initial Hilbert space,
with $S_{\mathsf{kk}}=\left[ S_{jk}\right] _{j,k\in \mathsf{k}}$ unitary,
and $H$ self-adjoint. (We use the convention that $L_{\mathsf{k}}=\left[
L_{k}\right] _{k\in \mathsf{k}}$ and that \ $L_{\mathsf{k}}^{\ast }L_{%
\mathsf{k}}=\sum_{k\in \mathsf{k}}L_{k}^{\ast }L_{k}$.) The triple $\left(
S,L,H\right) $ are termed the \textit{Hudson-Parthasarathy parameters} of the open
system evolution. Explicitly, we have the relations
\begin{eqnarray}
H &=&E_{00}+\frac{1}{2} E_{0\mathsf{k}}  \mathrm{Im}  \bigg\{ \frac{I_{\mathsf{k}}  }{ I_{\mathsf{k}%
}+\frac{i}{2}E_{\mathsf{kk}}  } \bigg\} E_{\mathsf{k}0} ,
\label{eq:H from_E} \\
L_{\mathsf{k}}&=& -i\left( I_{%
\mathsf{k}}+\frac{i}{2}E_{\mathsf{kk}}\right) ^{-1}E_{\mathsf{k}0},
\label{eq:L from_E}
\\
S_{\mathsf{kk}} &=&\frac{I_{\mathsf{k}}-\frac{i}{2}E_{\mathsf{kk}}}{I_{%
\mathsf{k}}+\frac{i}{2}E_{\mathsf{kk}}},
\label{eq:S from_E}
\end{eqnarray}
where Im$\,X$ means $\frac{1}{2i}\left( X-X^{\ast }\right) $.

We refer to this  model as an SLH model, with $(S,L,H)$ being the Hudson-Parthasarathy parameters.
We include closed systems, determined by a Hamiltonian operator $H$ within this category, and give the designation $(- , -, H)$.

\begin{remark}
The matrix $S_{\mathsf{kk}}$ is a Cayley transformation of $E_{\mathsf{kk}}$%
, and the self-adjointness of $E_{\mathsf{kk}}$ implies the unitarity of $S_{%
\mathsf{kk}}$. Note that we may invert to get
\begin{eqnarray}
E_{\mathsf{kk}}=\frac{2}{i}\frac{I_{\mathsf{k}}-S_{\mathsf{kk}}}{I_{\mathsf{k%
}}+S_{\mathsf{kk}}},
\end{eqnarray}
provided that $I_{\mathsf{k}}+S_{\mathsf{kk}}$ is invertible. This latter
condition is not always satisfied so we note that there are exceptional SLH
models that do not possess Stratonovich representations. A simple example is
an optical mirror for which $S\equiv -1$, which may only be obtained as a singular
limit $\lim_{\varepsilon \rightarrow \infty }\frac{1-\frac{i}{2}\varepsilon
}{1+\frac{i}{2}\varepsilon }$.
\end{remark}

We look at some special cases.

\subsubsection{$LH$ Models}

Let us first look at the case where $E_{\mathsf{kk}}=0$. Here the quantum
stochastic Hamiltonian is
\begin{eqnarray*}
\Upsilon \left( t\right) =E_{00}+\sum_{j\in \mathsf{k}}E_{j0}b_{j}\left(
t\right) ^{\ast }+\sum_{k\in \mathsf{k}}E_{0k}\ b_{k}\left( t\right) .
\end{eqnarray*}
This describes a standard emission-absorption interaction for the input
field quanta. \ For the QSDE, we simply have
\begin{gather}
dU\left( t\right) =\bigg\{-\left( \frac{1}{2}L_{\mathsf{k}}^{\ast }L_{%
\mathsf{k}}+iH\right) \otimes dt  \nonumber \\
+\sum_{j\in \mathsf{k}}L_{j}\otimes dB_{j}\left( t\right) ^{\ast
}-\sum_{j\in \mathsf{k}}L_{j}^{\ast }\otimes dB_{j}\left( t\right) \bigg\}%
U\left( t\right) .
\end{gather}
We note that in this case $S_{\mathsf{kk}}=I_{\mathsf{k}}$. We have $H = E_{00}$ and $L_{\mathsf{k}}= -i E_{\mathsf{k}0}$.

\subsubsection{$S$ Models}

If we set $E_{00}=E_{0\mathsf{k}}=E_{\mathsf{k}0}=0$, then
\begin{eqnarray*}
\Upsilon \left( t\right) =\sum_{j,k\in \mathsf{k}}E_{jk}b_{j}\left( t\right)
^{\ast }b_{k}\left( t\right) ,
\end{eqnarray*}
and the QSDE is

\begin{eqnarray}
dU\left( t\right) =\sum_{j,k\in \mathsf{k}}(S_{jk}-\delta _{jk})\otimes
d\Lambda _{jk}\left( t\right) \ U\left( t\right)
\end{eqnarray}
In this case, we have pure scattering only, with the scattering operator matrix given by
(\ref{eq:S from_E}).

\subsection{Network Composition Rules}

\subsubsection{The Series Product}
\begin{figure}[tbph]
\centering
\includegraphics[width=0.30\textwidth]{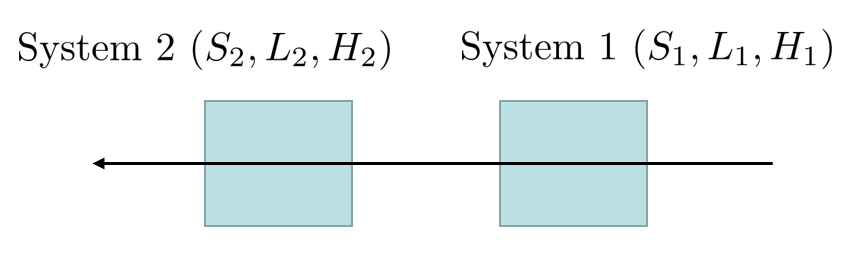}
\caption{Systems in series: Output of system 1 is the input of system 2.}
\label{fig:SS_Series}
\end{figure}

In \cite{GJ_Series}, we introduced the series product describing the
situation where one SLH drives another, see Figure \ref{fig:SS_Series}. Here
the output of the first system $\left( S_{1},L_{1},H_{1}\right) $ is fed
forward as the input to the second system $\left( S_{2},L_{2},H_{2}\right) $
and the limit of zero time delay is assumed. (Note that the systems do not
technically have to be distinct and may have the same initial space!) In the
limit, the Hudson-Parthasarathy parameters of the composite system were
shown to be \cite{GJ_Series,GJ_QFN}
\begin{eqnarray*}
S_{\mathrm{series}} &=&S_{2}S_{1}, \\
L_{\mathrm{series}} &=&L_{2}+S_{2}L_{1}, \\
H_{\mathrm{series}} &=&H_{1}+H_{2}+\mathrm{Im}\left\{ L_{2}^{\dag
}S_{2}L_{1}\right\} .  \label{eq:series_product}
\end{eqnarray*}

We refer to the associative group law
\begin{eqnarray*}
&&\left( S_{2},L_{2},H_{2}\right) \vartriangleleft \left(
S_{1},L_{1},H_{1}\right)  \notag \\
&&= \left( S_{2}S_{1},L_{2}+S_{2}L_{1},H_{1}+H_{2}+\mathrm{Im}\{L_{2}^{\dag
}S_{2}L_{1}\}\right)
\end{eqnarray*}
determined above, as the series product.

\subsubsection{Concatenation}
We also have a rule for concatenating separate models $\left( S_{j},L_{j},H_{j}\right) _{j=1}^{n}$ in parallel,  so as to obtain a single SLH component \cite{GJ_QFN}
\begin{eqnarray}  \label{eq:SLH_parallel}
&&\boxplus _{j=1}^{n}\left( S_{j},L_{j},H_{j}\right) =  \notag \\
&& \left( \left[
\begin{array}{ccc}
S_{1} & 0 & 0 \\
0 & \ddots & 0 \\
0 & 0 & S_{n}
\end{array}
\right] ,\left[
\begin{array}{c}
L_{1} \\
\vdots \\
L_{n}
\end{array}
\right] , H_{1}+\cdots +H_{n}\right) .  \notag \\
\end{eqnarray}

(Note that we have made no assumptions that the operators corresponding to different components commute!)

\subsubsection{Feedback}

\begin{figure}[h]
\centering
\includegraphics[width=0.4\textwidth]{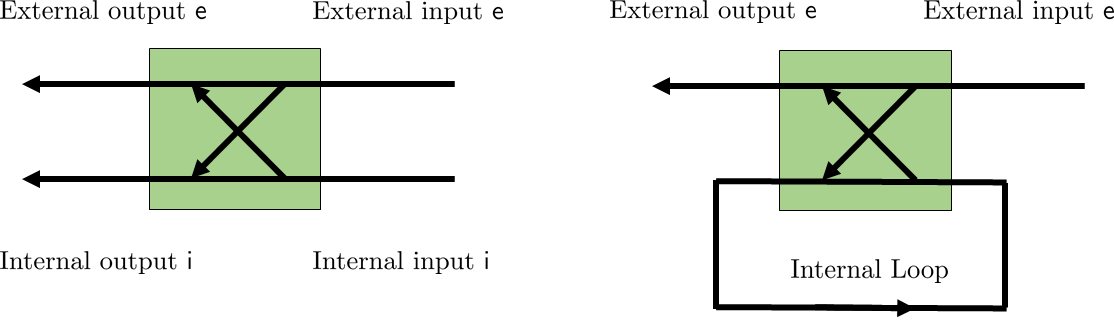}
\caption{(color online) Left: A quantum Markov model. Right: A feedback loop model.}
\label{fig:SLH_feedback_reduction}
\end{figure}

We now consider the situation where we take our original set of inputs and
outputs, labeled by $\mathsf{k}$, and select a certain subset of outputs to
be fed back back in as inputs. Without loss of generality we may assume that
the labels of these fields match and denote the index set as $\mathsf{i}$
(the internal indices). The remaining set $\mathsf{e}$ label the
external fields. See Figure \ref{fig:SLH_feedback_reduction}.

It follows that $\mathsf{k}$ is the disjoint union $\mathsf{e}\cup \mathsf{i}
$. The Hamiltonian $\Upsilon \left( t\right) $ may then be decomposed as
\begin{eqnarray*}
\left.
\begin{array}{c}
\left[ 1,b_{\mathsf{e}}\left( t\right)^{\dag} b_{\mathsf{i}}\left( t\right)^{\dag} \right]  \\
\,
\end{array}
\right. \left[
\begin{array}{ccc}
E_{00} & E_{0\mathsf{e}} & E_{0\mathsf{i}} \\
E_{\mathsf{e}0} & E_{\mathsf{ee}} & E_{\mathsf{ei}} \\
E_{\mathsf{i}0} & E_{\mathsf{ie}} & E_{\mathsf{ii}}
\end{array}
\right] \left[
\begin{array}{c}
1 \\
b_{\mathsf{e}}\left( t\right)  \\
b_{\mathsf{i}}\left( t\right)
\end{array}
\right] .
\end{eqnarray*}
Applying the feedback connections we should obtain a reduced model where the
internal inputs have been eliminated leaving only the set $\mathsf{e}$ of
external fields. That is, we should obtain a stochastic Hamiltonian of the
form
\begin{eqnarray}
\Upsilon ^{\mathrm{fb}}\left( t\right) =\left.
\begin{array}{c}
\left[ 1,b_{\mathsf{e}}\left( t\right) ^{\dag }\right]  \\
\quad
\end{array}
\right. \mathbf{E}^{\mathrm{fb}}\left[
\begin{array}{c}
1 \\
b_{\mathsf{e}}\left( t\right)
\end{array}
\right] .
\end{eqnarray}
It is a remarkable result that the feedback
reduction formula for the Stratonovich form is actually the Schur complement
of the matrix $\mathbf{E}$ describing the open loop network where we short
out the internal blocks, \cite{G_StratQFN}:
\begin{eqnarray}
\mathbf{E}^{\mathrm{fb}} &\equiv& \left[
\begin{array}{cc}
E_{00}^{\mathrm{fb}} & E_{0\mathsf{e}}^{\mathrm{fb}} \\
E_{\mathsf{e}0}^{\mathrm{fb}} & E_{\mathsf{ee}}^{\mathrm{fb}}
\end{array}
\right] \nonumber \\
&=& \left[
\begin{array}{cc}
E_{00} & E_{0\mathsf{e}} \\
E_{\mathsf{e}0} & E_{\mathsf{ee}}
\end{array}
\right] -\left[
\begin{array}{c}
E_{0\mathsf{i}} \\
E_{\mathsf{ei}}
\end{array}
\right] E_{\mathsf{ii}}^{-1}\left[
\begin{array}{cc}
E_{\mathsf{i}0} & E_{\mathsf{ie}}
\end{array}
\right]  \nonumber \\
&=& \underset{\mathsf{i}}{\mathrm{Schur}} \,  \mathbf{E} .
\label{eq:SchurE}
\end{eqnarray}
The invertibility of the matrix $E_{\mathsf{ii}}$ of operators is
equivalent to the condition for well-posedness of the network.

The expression for the reduced coefficients for the It\={o}
form has been derived  in \cite{GJ_QFN}. Accordingly to the partition of inputs and outputs, we have a
block partition of the system parameters:
\begin{eqnarray*}
S_{\mathsf{kk}}=\left[
\begin{array}{cc}
S_{\mathsf{ee}} & S_{\mathsf{ei}} \\
S_{\mathsf{ie}} & S_{\mathsf{ii}}
\end{array}
\right] ,\,L_{\mathsf{k}}=\left[
\begin{array}{c}
L_{\mathsf{e}} \\
L_{\mathsf{i}}
\end{array}
\right] .
\end{eqnarray*}
The feedback system parameters are then
\begin{eqnarray}
S^{\mathrm{fb}} &=&S_{\mathsf{ee}}+S_{\mathsf{ei}}\left( I_{\mathsf{i}}-S_{%
\mathsf{ii}}\right) ^{-1}S_{\mathsf{ie}},  \label{eq:S_feedback} \\
L^{\mathrm{fb}} &=&L_{\mathsf{e}}+S_{\mathsf{ei}}\left( I_{\mathsf{i}}-S_{%
\mathsf{ii}}\right) ^{-1}L_{\mathsf{i}},  \label{eq:L_feedback} \\
H^{\mathrm{fb}} &=&H+L_{\mathsf{i}}^{\dag }\mathrm{Im} \bigg\{ S_{\mathsf{ii}}\left(
I_{\mathsf{i}}-S_{\mathsf{ii}}\right) ^{-1} \bigg\} L_{\mathsf{i}}  \nonumber \\
&+&\mathrm{Im}L_{\mathsf{e}}^{\dag }S_{\mathsf{ei}}\left( I_{\mathsf{i}}-S_{%
\mathsf{ii}}\right) ^{-1}L_{\mathsf{i}}.  \label{eq:H_feedback}
\end{eqnarray}

We also note that we need the condition that the inverse of $I_{\mathsf{i}%
}-S_{\mathsf{ii}}$ exists: this is the condition needed to ensure that the
network is well-posed.

\begin{remark}
\label{rem:Z_i}
Let us introduce
\begin{eqnarray}
 Z_{\mathsf{i}} \triangleq  \mathrm{Im} \bigg\{ S_{\mathsf{ii}}\left(
I_{\mathsf{i}}-S_{\mathsf{ii}}\right) ^{-1} \bigg\} ,
\label{eq:Z_i}
\end{eqnarray}
which appears in (\ref{eq:S_feedback}).
It is easy to show that
\begin{eqnarray}
 Z_{\mathsf{i}} =  \mathrm{Im} \big\{  \left(
I_{\mathsf{i}}-S_{\mathsf{ii}}\right) ^{-1} \big\} .
\label{eq:Z_ii}
\end{eqnarray}
\end{remark}

\begin{remark}
We may include an additional unitary ``gain''  $\eta$ along the feedback loop, as in Figure \ref{fig:SS_feedback}.

\begin{figure}[htbp]
	\centering
		\includegraphics[width=0.40\textwidth]{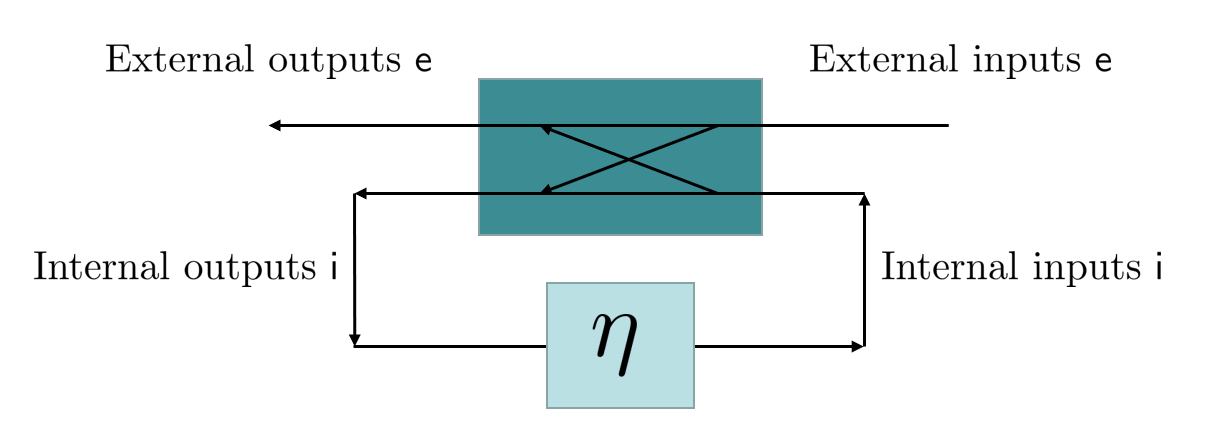}
	\caption{(color online) Feedback with a unitary gain matrix $\eta$.}
	\label{fig:SS_feedback}
\end{figure}

In this case the closed-loop terms are modified to
\begin{eqnarray}
S^{\mathrm{fb}} &=&S_{\mathsf{ee}}+S_{\mathsf{ei}}\left( \eta^{-1} -S_{\mathsf{ii}}\right) ^{-1}S_{\mathsf{ie}},    \nonumber\\
L^{\mathrm{fb}} &=&L_{\mathsf{e}}+S_{\mathsf{ei}}\left( \eta^{-1} -S_{\mathsf{ii}}\right) ^{-1}L_{\mathsf{i}},    \nonumber\\
H^{\mathrm{fb}} &=&H+L_{\mathsf{i}}^{\dag }\mathrm{Im} \bigg\{ S_{\mathsf{ii}}\left( \eta^{-1} -S_{\mathsf{ii}}\right) ^{-1} \bigg\} L_{\mathsf{i}}  \nonumber \\
&+&\mathrm{Im}L_{\mathsf{e}}^{\dag }S_{\mathsf{ei}}\left( \eta^{-1}-S_{\mathsf{ii}}\right) ^{-1}L_{\mathsf{i}}.  \label{eq:H_feedback_eta}
\end{eqnarray}

\end{remark}

\section{Isolated Loops in Quantum Networks}
\label{sec:Loops}

\begin{figure}[tbph]
\centering
\includegraphics[width=0.30\textwidth]{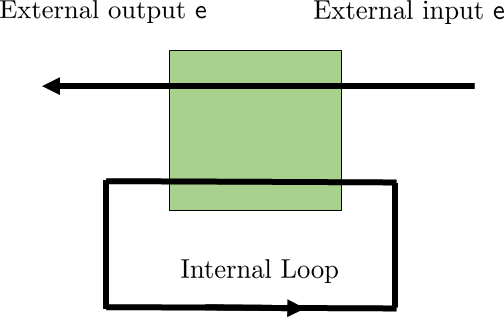}
\caption{(color online) A quantum feedback network with isolated algebraic
loops.}
\label{fig:SLH_internal_loop}
\end{figure}

The isolated loop case is given by setting $S_{\mathsf{ei}}=0$ and $S_{\mathsf{ie}}=0$, as in Figure \ref{fig:SLH_internal_loop}.

That is, $S$ is block-diagonal, and this happens if
and only if $E$ is block-diagonal:
\begin{eqnarray}
S_{\mathsf{kk}} \equiv \left[
\begin{array}{cc}
S_{\mathsf{ee}} & 0 \\
0 & S_{\mathsf{ii}}
\end{array}
\right] \Leftrightarrow
E_{\mathsf{kk}} \equiv \left[
\begin{array}{cc}
E_{\mathsf{ee}} & 0 \\
0 & E_{\mathsf{ii}}
\end{array}
\right] .
\end{eqnarray}

We now have $S_{\mathsf{ee}} \equiv \frac{I_{\mathsf{e}}-\frac{i}{2}E_{%
\mathsf{ee}}}{I_{\mathsf{e}}+\frac{i}{2}E_{\mathsf{ee}}}$ and
\begin{eqnarray}
S_{\mathsf{ii}} &\equiv & \frac{I_{\mathsf{i}}-\frac{i}{2}E_{\mathsf{ii}}}{%
I_{\mathsf{i}}+\frac{i}{2}E_{\mathsf{ii}}} , \label{eq:S_i}  \\
L_{\mathsf{i}} &\equiv &i\frac{I_{\mathsf{i}}}{I_{\mathsf{i}}+\frac{i}{2}E_{%
\mathsf{ii}}}E_{\mathsf{i}0} .  \label{eq:L_i}
\end{eqnarray}
Once again the well-posedness condition is equivalent to invertibility of
the operator matrices $I_{\mathsf{e}}-S_{\mathsf{ee}}$ and $I_{\mathsf{i}}-S_{\mathsf{ii}}$.

After feedback reduction, we have $S^{\text{fb}}=S_{\mathsf{ee}}$, $L^{\text{fb}}=L_{\mathsf{e}}$ and
\begin{eqnarray}
H^{\text{fb}}=H+V_{\mathrm{loop}} ,  \label{eq:H_loop}
\end{eqnarray}
where the internal loop Hamiltonian is
\begin{eqnarray}
V_{\mathrm{loop}} = \text{Im}\thinspace \left\{ L_{\mathsf{i}}^{\dag }S_{%
\mathsf{ii}}\left( I_{\mathsf{i}}-S_{\mathsf{ii}}\right) ^{-1}L_{\mathsf{i}}
\right\} .  \label{eq:V_loop}
\end{eqnarray}
Using Remark \ref{rem:Z_i}, we may write this more compactly as
\begin{eqnarray}
V_{\mathrm{loop}} = L_{\mathsf{i}}^\dag \, \text{Im}\thinspace \left\{  \left(
I_{\mathsf{i}}-S_{\mathsf{ii}}\right) ^{-1} \right\} L_{\mathsf{i}} \equiv
L_{\mathsf{i}}^\dag \, Z_{\mathsf{i}}  L_{\mathsf{i}} .
\label{eq:V_loop_simple}
\end{eqnarray}
We therefore have that
\begin{eqnarray*}
\left( S^{\text{fb}},L^{\text{fb}},H^{\text{fb}}\right) = \left( S_{\mathsf{%
ee}},L_{\mathsf{e}},H \right) \boxplus \left( \_,\_,V_{\mathrm{loop}}
\right) .
\end{eqnarray*}

\begin{remark}
In the present case, $S_{\mathsf{ii}}$ is unitary and related to $E_{\mathsf{ii}}$ by (\ref{eq:S_i}). Substituting in,
we see that
\begin{eqnarray}
Z_{\mathsf{i}} \equiv - E_{\mathsf{ii}}^{-1}.
\end{eqnarray}
Moreover, using the unitarity of $S_{\mathsf{ii}}$, we find from (\ref{eq:Z_ii})
\begin{eqnarray}
Z_{\mathsf{i}} \equiv \frac{1}{2i} \frac{I_{\mathsf{i}} +S_{\mathsf{ii}} }{ I_{\mathsf{i}} - S_{\mathsf{ii}} }.
\label{eq:Z_iii}
\end{eqnarray}
As we have commented above, the invertibility of $E_{\mathsf{ii}}$ is equivalent to the
well-posedness of the feedback loop. In the isolated loop case it also implies the existence of the self-adjoint $Z_{\mathsf{i}}$.
\end{remark}

\begin{proposition}
Making the assumption that the internal loop is well-posed, then the
feedback reduced Hamiltonian $H^{\mathrm{fb}}=H+ V_{\mathrm{loop}}$ may be
written as
\begin{eqnarray}
H^{\mathrm{fb}} = \underset{\mathsf{i}}{ \mathrm{Schur}} \left[
\begin{array}{cc}
E_{00} & E_{0\mathsf{i}} \\
E_{\mathsf{i}0} & E_{\mathsf{ii}}
\end{array}
\right]   + V_{\mathsf{e}} ,  \label{eq:prop}
\end{eqnarray}
where $V_{\mathsf{e}} \triangleq \frac{1}{2}E_{0\mathsf{e}}\,\text{Im}%
\,\left\{ \frac{I_{\mathsf{e}}}{I_{\mathsf{e}}+\frac{i}{2}E_{\mathsf{ee}}}%
\right\} E_{\mathsf{e}0}$.
\end{proposition}
Note that we may equivalently write (\ref{eq:prop}) as
\begin{eqnarray}
H^{\mathrm{fb}} &=& E_{00}-E_{0\mathsf{i}}\left( E_{\mathsf{ii}}\right) ^{-1}E_{%
\mathsf{i}0} + V_{\mathsf{e}} \nonumber\\
&=& E_{00}+ E_{0\mathsf{i}} Z_{\mathsf{i}} E_{\mathsf{i}0} + V_{\mathsf{e}} .
\end{eqnarray}

\begin{proof}
The well-posed property is equivalent to the invertibility of $E_{\mathsf{ii}%
}$ and therefore ensures the existence of the Schur complement. As $E$ is
block-diagonal, we have that the expression for $H$ in (\ref{eq:H from_E})
reduces to
\begin{eqnarray*}
E_{00}+ \frac{1}{2}\sum_{\mathsf{a} = \mathsf{e}, \mathsf{i} } \mathrm{Im}
\left\{ E_{0\mathsf{a}} \left( I_{\mathsf{a}}+\frac{i}{2}E_{\mathsf{aa}%
}\right) ^{-1} E_{\mathsf{a}0} \right\} .
\end{eqnarray*}
Combining this with (\ref{eq:H_loop}), we obtain
\begin{eqnarray*}
H^{\text{fb}} &=& H + V _{\mathrm{loop}} \\
&=&E_{00}+ V_{\mathsf{e}} +\frac{1}{2}E_{0\mathsf{i}}\,\text{Im}\,\left\{
\frac{I_{\mathsf{i}}}{I_{\mathsf{i}} +\frac{i}{2}E_{\mathsf{ii}}}\right\} E_{%
\mathsf{i}0} \\
&& +\text{Im }\,\left\{ L_{\mathsf{i}}^{\dag }S_{\mathsf{ii}}\left( I_{%
\mathsf{i}}-S_{\mathsf{ii}}\right) ^{-1}L_{\mathsf{i}}\right\} ,
\end{eqnarray*}
and by direct substitution of (\ref{eq:L_i}), we have that
\begin{eqnarray*}
L_{\mathsf{i}}^{\dag }S_{\mathsf{ii}}\left( I_{\mathsf{i}}-S_{\mathsf{ii}%
}\right) ^{-1}L_{\mathsf{i}} =E_{0\mathsf{i}}\left\{ \frac{I_{\mathsf{i}}}{%
I_{\mathsf{i}}+\frac{i}{2}E_{\mathsf{ii}}}\frac{I_{\mathsf{i}}}{\frac{i}{2}%
E_{\mathsf{ii}}}\right\} E_{\mathsf{i}0}
\end{eqnarray*}
and so
\begin{eqnarray*}
H^{\text{fb}}=E_{00}+\frac{1}{2}E_{0\mathsf{i}}\text{Im}\,\left\{ \frac{I_{%
\mathsf{i}}}{\frac{i}{2}E_{\mathsf{ii}}}\right\} E_{\mathsf{i}0} + V_{%
\mathsf{e}},
\end{eqnarray*}
which gives the desired form (\ref{eq:prop}).
\end{proof}

\subsection{Completely Isolated Loops}
\label{subsec:Completely Isolated Loops}

\begin{figure}[htbp]
	\centering
		\includegraphics[width=0.20\textwidth]{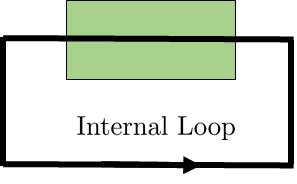}
	\caption{(color online) A completely isolated system.}
	\label{fig:Single_Loop}
\end{figure}

We consider an SLH system where we create a completely isolated loop by feeding all of the
outputs back as inputs; see Figure \ref{fig:Single_Loop}. For convenience, we drop the label indices. The feedback reduction formula suggests that the resulting system is a closed Hamiltonian system, and using (\ref{eq:H_loop})
and (\ref{eq:V_loop_simple}), the Hamiltonian will be
\begin{eqnarray}
H_{\text{loop}} &=& H+\text{Im}\left\{ L^{\ast }\frac{1}{1-S}L\right\} \nonumber \\
&=& H + L^\dag Z \, L.
\label{eq:H_loops}
\end{eqnarray}
Here $Z= \frac{1}{2i} \frac{I+S}{I-S}$, as in (\ref{eq:Z_iii}). We see that
$H_{\text{loop}} $ is indeed a self-adjoint operator provided $I-S$ is invertible (the well-posedness criterion for the feedback loop).

\begin{figure}[htbp]
\centering
\includegraphics[width=0.40\textwidth]{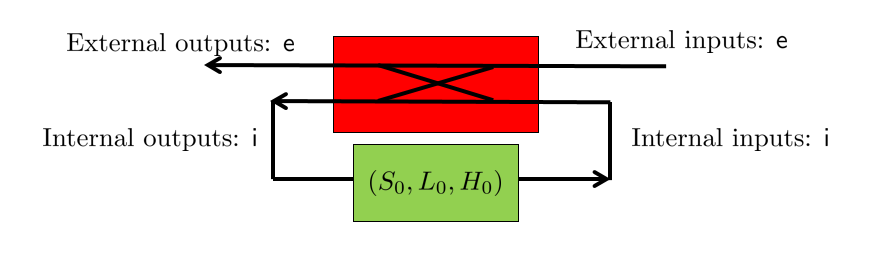}
\caption{(color online) A component $\left( S_{0},L_{0},H_{0}\right) $ in a
well-defined algebraic loop.}
\label{fig:SLH_limit_loop}
\end{figure}

We re-derive (\ref{eq:H_loops}) by considering the limit of a regular network with external inputs and outputs, leading to a completely isolated loop. This is represented in Figure \ref{fig:SLH_limit_loop} with an arbitrary component $\left(S_{0},L_{0},H_{0}\right) $ in loop.

The top component is a simple beam-splitter with a transmission coefficient $%
t$. Before making the feedback connections we have the open loop SLH model
with
\begin{eqnarray*}
S =\left[
\begin{array}{cc}
S_{\mathsf{ee}} & S_{\mathsf{ei}} \\
S_{\mathsf{ie}} & S_{\mathsf{ii}}
\end{array}
\right] ,\, L =\left[
\begin{array}{c}
L_{\mathsf{e}} \\
L_{\mathsf{i}}
\end{array}
\right] ,\, H =H_{0},
\end{eqnarray*}
where
\begin{align*}
S_{\mathsf{ee}} &= \sqrt{1-t^{2}} I_n, & 
S_{\mathsf{ei}} &=\left[-t I_n \ 0\right] ,\\
S_{\mathsf{ie}} &=\left[
\begin{array}{c}
t I_n \\
0
\end{array}
\right], & S_{\mathsf{ii}} &=\left[
\begin{array}{cc}
\sqrt{1-t^{2}} I_n & 0 \\
0 & S_{0}
\end{array}
\right], \\
L_{\mathsf{e}} &=0, &
L_{\mathsf{i}} &=\left[
\begin{array}{c}
0 \\
L_{0}
\end{array}
\right].
\end{align*}
$\mathsf{e}$ labels the $n$ external channels,
and $\mathsf{i}\equiv \mathsf{i}_{1}\cup \mathsf{i}_{2}$ where $\mathsf{i}%
_{1}$ are the remaining inputs into the beam-splitter and $\mathsf{i}_{2}$
the inputs into the component $\left( S_{0},L_{0},H_{0}\right) $.
Also. the adjacency matrix (unitary gain)
$\eta= \bigl[\begin{smallmatrix} 0 & I_n \\ I_n & 0 \end{smallmatrix}\bigr]$
is used in the feedback loop. Substituting into (\ref{eq:S_feedback}) leads to
\begin{eqnarray*}
S\left( t\right) &=&\left( 1-t^{2}\right) I_{n}+t^{2}\left( S_{0}^{-1}-\sqrt{%
1-t^{2}}I_{n}\right) ^{-1}, \\
L\left( t\right) &=&-t\left( S_{0}^{-1}-\sqrt{1-t^{2}}I_{n}\right)
^{-1}L_{0}, \\
H\left( t\right) &=&H_{0}
+\sqrt{1-t^{2}}\,\text{Im}\left\{\! L_{0}^{\dag }\left( I_{n}-\sqrt{1-t^{2}}%
S_{0}\right) ^{-1}\! L_{0}\!\right\} .
\end{eqnarray*}
The limit $t\rightarrow 0$ is well-defined and leads to
\begin{eqnarray*}
S\left( 0\right) &=&I_{n},\quad L\left( 0\right) =0, \\
H\left( 0\right) &=&H_{0}+\text{Im} \left\{ L_{0}^{\dag
}(I_{n}-S_{0})^{-1}L_{0}\right\},
\end{eqnarray*}
This we recognize as the SLH model where the input is trivially reflected -
and so equals the output - while the Hamiltonian is modified exactly as in (%
\ref{eq:H_loops}).

\subsection{The Series Product in an Isolated Loop}

For two components $\left( S_{1},L_{1},H_{1}\right) $ and $\left(
S_{2},L_{2},H_{2}\right) $ in series, we have the effective SLH component
given by (\ref{eq:series_product}). Note that the first system drives the
second and so the order is important.

\begin{figure}[h]
\centering
\includegraphics[width=0.40\textwidth]{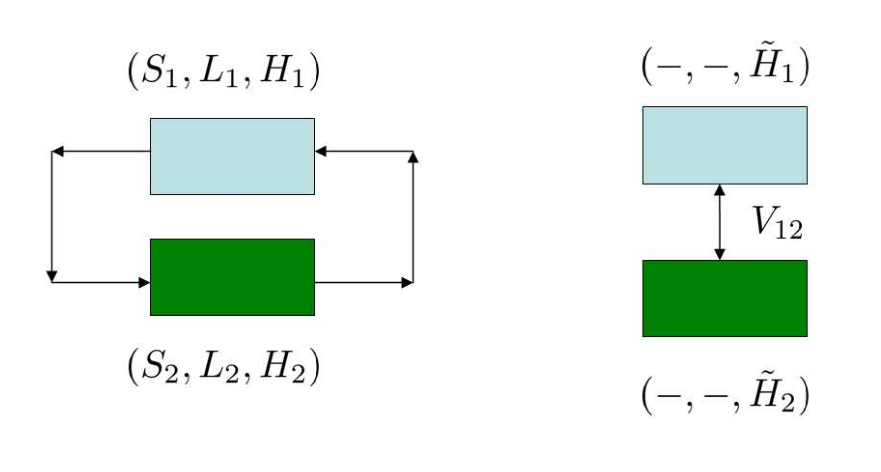}
\caption{(color online) An isolated loop connecting two SLH components}
\label{fig:SLH_series_loop}
\end{figure}

Connecting them into an isolated loop, as in Figure \ref{fig:SLH_series_loop},
leads to the effective model $\left( \_,\_,H_{\text{loop}}\right) $ with
\begin{multline*}
H_{\text{loop}}=H_{1}+H_{2}+\text{Im }\,\left\{ L_{2}^{\dag
}S_{2}L_{1}\right\} \\
+\text{Im\thinspace }\left\{ (L_{2}+S_{2}L_{1})^{\dag }S_{2}S_{1}\frac{1}{%
I-S_{2}S_{1}}(L_{2}+S_{2}L_{1})\right\}
\end{multline*}
and the construction is well-posed so long as $I-S_{2}S_{1}$ is invertible.

One would naturally expect this Hamiltonian to be symmetric under
interchange of the systems 1 and 2, and indeed it is not difficult to see
that we may rearrange to get
\begin{eqnarray*}
H_{\text{loop}} &=&H_{1}+\text{Im}\,\left\{ L_{1}^{\dag }S_{1}\frac{1}{%
I-S_{2}S_{1}}(L_{2}+S_{2}L_{1})\right\} \\
&+&H_{2}+\text{Im}\,\left\{ L_{2}^{\dag }S_{2}\frac{1}{I-S_{1}S_{2}}%
(L_{1}+S_{1}L_{2})\right\} .
\end{eqnarray*}
We may also write this as
\begin{eqnarray*}
H_{\text{loop}}=\tilde{H}_{1}+\tilde{H}_{2}+V_{12}
\end{eqnarray*}
where the modified Hamiltonians of the components are
\begin{eqnarray*}
\tilde{H}_{1} &=&H_{1}+\text{Im}\,\left\{ L_{1}^{\dag }\frac{1}{I-S_{1}S_{2}}%
L_{1}\right\} , \\
\tilde{H}_{2} &=&H_{2}+\text{Im}\,\left\{ L_{2}^{\dag }\frac{1}{I-S_{2}S_{1}}%
L_{2}\right\} ,
\end{eqnarray*}
and we obtain an effective inter-component coupling
\begin{eqnarray}
V_{12} &=&\text{Im}\,\left\{ L_{1}^{\dag }S_{1}\frac{1}{I-S_{2}S_{1}}%
L_{2}\right\}  \nonumber \\
&&+\text{Im}\,\left\{ L_{2}^{\dag }S_{2}\frac{1}{I-S_{1}S_{2}}L_{1}\right\} .
\label{eq:V_coupling}
\end{eqnarray}

\subsection{Modeling General Interactions}

Suppose we wished to obtain a specific interaction $V_{12}$ between a pair
of systems by setting up a loop. Let us suppose that we have
\begin{eqnarray*}
L_{1} &=&A,\qquad S_{1}=e^{i\phi _{1}}, \\
L_{1} &=&B,\qquad S_{2}=e^{i\phi _{2}}.
\end{eqnarray*}
Substituting into (\ref{eq:V_coupling}), we find
\begin{eqnarray}
V_{12}\equiv \lambda \, A\otimes B
\end{eqnarray}
where
\begin{eqnarray}
\lambda =\frac{\sin \phi _{1}+\sin \phi _{2}}{1-\cos \left( \phi _{1}+\phi
_{2}\right) }.
\end{eqnarray}
In principle, we may set $\lambda $ to any desired value by adjusting $\phi
_{1}$ and $\phi _{2}$. However, it is convenient to think of this as just a
constant.

More generally we can have multiple loops with
\begin{eqnarray*}
L_{1}=\left[
\begin{array}{c}
A_{1} \\
\vdots \\
A_{n}
\end{array}
\right] ,\quad L_{2}=\left[
\begin{array}{c}
B_{1} \\
\vdots \\
B_{n}
\end{array}
\right] ,
\end{eqnarray*}
and specified phases, so that
\begin{eqnarray*}
V_{12}\equiv \sum_{k}\lambda _{k}\,A_{k}\otimes B_{k}.
\end{eqnarray*}
Theoretically we can approximate any interaction coupling.

\section{Examples}
\label{sec:Examples}

\subsection{Completely Isolated Looped Cavity}

\begin{figure}[htbp]
\centering
\includegraphics[width=0.40\textwidth]{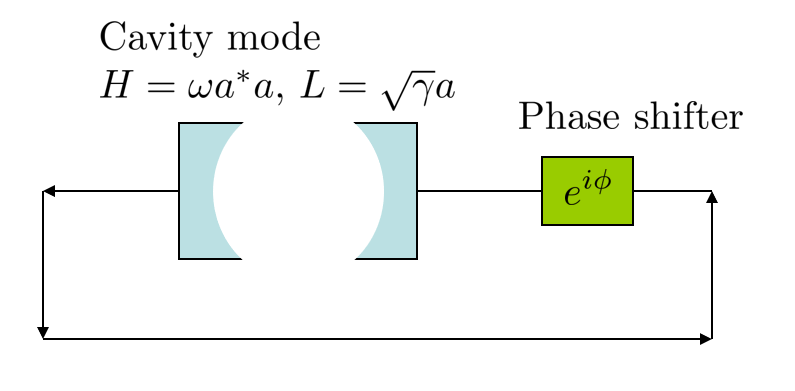}
\caption{(color online) Cavity loop}
\label{fig:cavity_loop}
\end{figure}

In the special case of a cavity mode $a$ with frequency $\omega $ and
damping rate $\gamma $ in series with a phase shifter $\left( S=e^{i\phi
}\right) $, see Figure \ref{fig:cavity_loop}, we find that
\begin{eqnarray*}
H_{\text{loop}}=\tilde{\omega}a^{\ast }a
\end{eqnarray*}
where the shifted frequency is
\begin{eqnarray*}
\tilde{\omega} &=&\omega +\gamma \, \text{Im}\frac{1}{1-e^{i\phi }} = \omega
+ \frac{\gamma }{2} \frac{\sin \phi }{ 1-\cos \phi }.
\end{eqnarray*}

The network is not well-posed when $\cos \phi =1$, so in particular we need
a nontrivial phase shift $e^{i \phi} \neq 1$ in the loop. We see that the
shifted frequency $\tilde{\omega}\left( \phi \right)$ can take any real value
as $\phi$ varies between $0$ and $2\pi$. In effect, we have created a closed cavity
with tunable resonant frequency.

%
%
%

\subsection{Two Cavities Interacting}

\begin{figure}[htbp]
\centering
\includegraphics[width=0.40\textwidth]{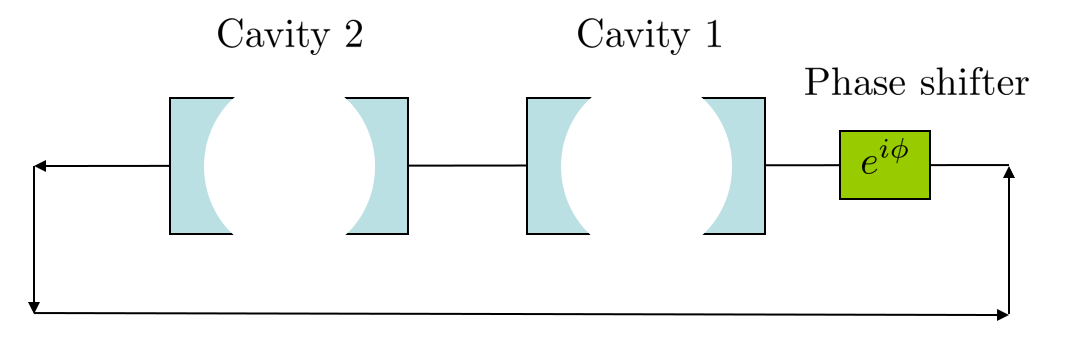}
\caption{(color online) A coupled cavity scheme arising from looped cavities
in series.}
\label{fig:SLH_cavity_loop2}
\end{figure}

Let us take two cavities with the SLH specifications $(S_1 =
e^{i\phi _{1}}, L_1 =\sqrt{\gamma _{1}}a_{1}, H_1 =\omega _{1}a_{1}^{\ast }a_{1} ) $%
, and $ ( S_2 = e^{i\phi _{2}}, L_2 = \sqrt{\gamma _{2}}a_{2}, H_2 =\omega
_{2}a_{2}^{\ast }a_{2} ) $.

The closed loop of Figure \ref{fig:SLH_cavity_loop2}, will then have shifted
frequencies for the two cavities, along with an additional interaction
\begin{eqnarray*}
V_{12}=\kappa _{12}\,a_{1}^{\ast }\otimes a_{2}+\kappa _{12}^{\ast}\,a_{2}^{\ast}\otimes a_{1},
\end{eqnarray*}
where
\begin{eqnarray*}
\kappa _{12}=\frac{\sqrt{\gamma _{1}\gamma _{2}}}{2i}f\left( \phi _{1},\phi _{2}\right)
\end{eqnarray*}
and
\begin{eqnarray}
f\left( \phi _{1},\phi _{2}\right) =\frac{e^{i\phi _{1}}-e^{-i\phi _{2}}}{%
1-\cos \left( \phi _{1}+\phi _{2}\right) }.  \label{eq:f}
\end{eqnarray}

\subsection{Jaynes-Cummings Interaction}

Let us consider a two-level system with Hilbert space $\mathfrak{h}_{0}\cong
\mathbb{C}^{2}$ coupled to a cavity mode with Hilbert space $\mathfrak{h}_{%
\text{mode}}$. The most general interaction possible can be written as
\begin{eqnarray*}
V=V_{0}\otimes I_{2}+V_{+}\otimes \sigma _{-}+V_{-}\otimes \sigma
_{+}+V_{z}\otimes \sigma _{z}
\end{eqnarray*}
with respect to the factorization $\mathfrak{h}_{\text{mode}}\otimes \mathbb{%
C}^{2}$, and with $\sigma _{\pm },\sigma _{z}$ being the usual spin
matrices. Here $V_{0},V_{\pm }$ and $V_{z}$ are operators on the Hilbert
space $\mathfrak{h}_{\text{mode}}$ with $V_{0}^{\ast }=V_{0}$, $V_{\pm
}^{\ast }=V_{\mp }$ and $V_{z}^{\ast }=V_{z}$. We shall take $V_{0}\equiv 0$
as this does not specify a coupling between the two-level system and the
mode.

\begin{figure}[h]
\centering
\includegraphics[width=0.30\textwidth]{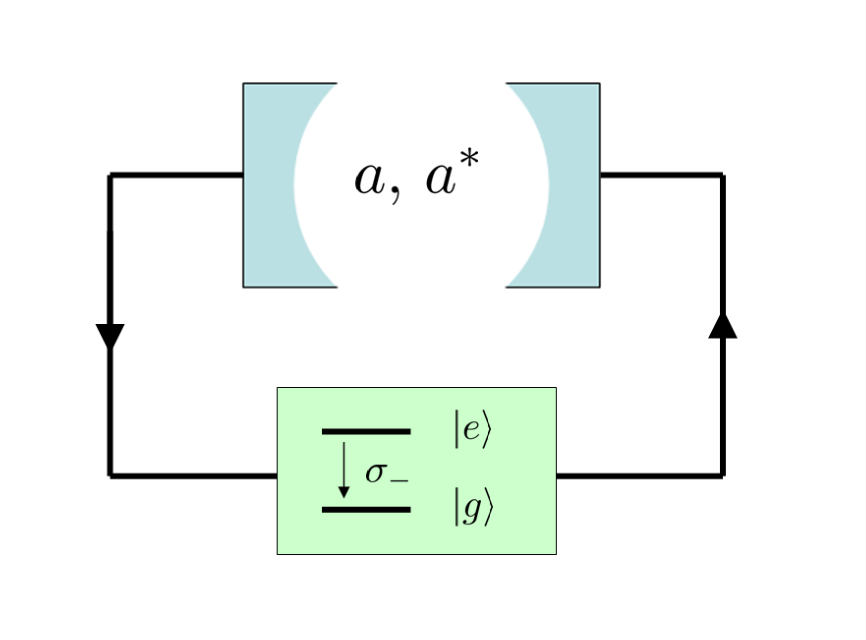}
\caption{(color online) A cavity in an isolated loop with a two level atom}
\label{fig:SLH_JC}
\end{figure}

In order to realize the interaction, we could consider the two loops (i.e.,
the multiplicity $n=2$ in the loop in Figure \ref{fig:SLH_JC}) with
\begin{eqnarray*}
L_{1}=L_{\text{mode}}=\left[
\begin{array}{c}
X \\
Y
\end{array}
\right] ,\quad L_{2}=\left[
\begin{array}{c}
\sqrt{\gamma }\sigma _{-} \\
\sqrt{\kappa }\sigma _{z}
\end{array}
\right] ,
\end{eqnarray*}
and scattering corresponding to phases $S_1 = e^{i\phi _{1}}I_2$ and $S_2 =
e^{i\phi _{2}}I_2$ respectively. From (\ref{eq:V_coupling}), this leads to
the interaction
\begin{eqnarray*}
V_{12}=V_{+}\otimes \sigma _{-}+V_{-}\otimes \sigma _{+}+V_{z}\otimes \sigma
_{z}
\end{eqnarray*}
with
\begin{eqnarray*}
V_{+} &=&\frac{\sqrt{\gamma }f\left( \phi _{1},\phi _{2}\right) }{2i}X, \\
V_{z} &=&\frac{\sqrt{\kappa }}{2i}\left( f\left( \phi _{1},\phi _{2}\right)
Y-f\left( \phi _{1},\phi _{2}\right) ^{\ast }Y^{\ast }\right)
\end{eqnarray*}
and $f\left( \phi _{1},\phi _{2}\right) $ as in (\ref{eq:f}) above.

In particular, this leads to a way to engineer a Jaynes-Cummings interaction
between the cavity mode and the two-level system with \textit{adjustable}
coupling constants.

\section{Isolated Loops with Time Delays}
\label{sec:Delay}
There has been recent interest in modeling time-delays in optical communication networks where
signals have to traverse sizeable distances. Trapped modes arise for instance in truncated models \cite{Tabak_Mabuchi},
however, we wish to discuss isolated modes as exact system models with a finite time-delay.

Let us consider an assembly of $m$ oscillators, with modes $a = [a_1 , \cdots , a_m ]^\top$, as plant with
Hamiltonian $H= a^{\dag} \Omega a$ (with $\Omega\in \mathbb{C}^{m\times m}$ a Hermitian matrix), coupling operators
$L = C a$ (with $C\in \mathbb{C}^{n \times m}$), and $S \in \mathbb{C}^{n \times n}$ a unitary scattering matrix.

The Heisenberg equations of motion lead to a linear dynamical system
\begin{eqnarray}
\dot{a} (t) &=& A \, a(t) + B \, b_{\mathrm{in}} (t) ,\nonumber \\
b_{\mathrm{out}}(t) &=& C \, a(t) + D \, b_{\mathrm{in}} (t),
\end{eqnarray}
where $A= - \frac{1}{2} C^{\dag} C - i \Omega$, $B=-C^{\dag} S$ and $D=S$.

We now consider closing the feedback loop, but with a time delay $\tau >0$:
\begin{eqnarray}
b_{\mathrm{in}} (t) = b_{\mathrm{out}}(t -\tau ) .
\end{eqnarray}

To proceed, we work in the Laplace domain and set $a[s] \triangleq \int_0^\infty e^{-st} a(t)dt$, etc.
This leads to
\begin{eqnarray}
a[s] +a(0) &=& A \, a[s] + B \, b_{\mathrm{in}} [s] ,\nonumber \\
b_{\mathrm{out}} [s] &=& C \, a[s] + D \, b_{\mathrm{in}} [s],
\end{eqnarray}
with the feedback condition being $b_{\mathrm{in}} [s] = e^{s \tau } b_{\mathrm{out}} [s]$.

Again, it is possible to eliminate the internal fields, and this leaves us with
\begin{eqnarray}
a[s] = \frac{1}{ sI- A^{\mathrm{fb}} (s) } a(0) ,
\end{eqnarray}
where we now have (compare (\ref{eq:H_feedback_eta}) with $\eta =e^{s\tau }$)
\begin{eqnarray}
A^{\mathrm{fb}} (s)  = A  + B \big( e^{-s \tau}I -D \big)^{-1} C  \equiv  -i \Omega_{\mathrm{fb}} (s),
\end{eqnarray}
where
\begin{eqnarray}
 \Omega_{\mathrm{fb}}(s)  \triangleq   \Omega -i C^\dag \bigg( \frac{1}{2} +S ( e^{-s \tau}I -S )^{-1} \bigg) C . \nonumber \\
\quad
\label{eq:Omega_fb}
\end{eqnarray}
We note that as $\tau \to 0$ we recover $\Omega_{\mathrm{fb}}= \Omega +\frac{1}{2i} C^\dag   \frac{I+S}{I-S} C$ in line with
(\ref{eq:H_loops}).

\section{Conclusions}
\label{sec:Conclusions}

We have shown that it is possible in quantum feedback networks to have feedback loops that are isolated in the sense that there are no optical
signal traveling between the degrees of freedom of the subsystem associated with the loop and
those of the rest of the network. In fact, these can arise in very natural ways. Despite the fact that the optical loop is closed, it is shown nevertheless
that generally there will be an effective (direct Hamiltonian) interaction set up between these components. We show that these interactions can be designed
by setting up appropriate interconnections - in a sense, a form of reservoir engineering where the reservoir is an isolated loop. The dependence of the
interaction on internal scattering parameters is derived, and may be used as tuning parameters for interactions. We also determine the dependence on
time delay for linear passive optical loops.

\section{Acknowledgements}

The paper was motivated by discussions at the workshop Quantum and
Nano-Control at the Institute of Mathematics and Its Applications in
Minnesota during April 2016, and the kind support and hospitality of the IMA
is greatly acknowledged. Also Support of the Australian Research Council
and the Air Force Office of Scientific Research is acknowledged.

\end{document}